\documentstyle[mncite,psfig]{mn}

\title[Resonance Doublets in Bipolar Winds]
{Profiles of the Resonance Doublets Formed in Bipolar Winds
in Symbiotic Stars}

\author[Yoo, Lee and Ahn]
{Jerry Jaiyul Yoo$^{1}$\thanks{E-mail: jyyu@astro.snu.ac.kr},
Hee-Won Lee$^2$ and Sang-Hyeon Ahn$^3$\\
$^{1}$Department of Astronomy,
Seoul National University, Shillim-Dong, Kwanak-Gu, Seoul, Korea\\
$^{2}$Department of Geoinformation Science, Sejong University, Gunja-Dong,
Gwangjin-Gu, Seoul, Korea\\
$^{3}$Korea Institute for Advanced Study, Cheongyangri-Dong,
Dongdaemun-Gu, Seoul, Korea}

\date{Accepted 2002 April 17. Received 2002 March 20; in original form
2002 January 5}

\begin{document}
\maketitle

\begin{abstract}
We compute the profiles of resonance doublet lines ($S_{1/2}-P_{1/2,3/2}$)
formed in bipolar
winds with velocity greater than the doublet separation in symbiotic stars.
Particular attention has been paid on the doublet line ratio, where
an essential role is played by the conversion of the short wavelength 
component arising from the $S_{1/2}-P_{3/2}$ transition into the long 
wavelength component for the transition $S_{1/2}-P_{1/2}$.
We adopted a Monte Carlo technique and
the Sobolev approximation. Our bipolar winds take the form of a cone and 
are characterized by the terminal wind velocity, the mass loss rate and 
the opening angle of the cone. When an observer is in the polar direction
and the Sobolev optical depth $\tau_{Sob}\simeq 1$, we mainly
obtain profiles with inverted flux line ratios, where the short wavelength
component is weaker than the long wavelength component. When an observer
is in the equatorial direction, we find that the profiles are characterized
by two broad components, where the long wavelength component is the broader and
stronger of the two. We conclude that the
profiles obtained in our model provide a qualitative understanding
of broad profiles and inverted intensity ratios of the doublets 
in symbiotic stars.

\end{abstract}


\section{Introduction}
Symbiotic stars are generally known to be interacting binaries
consisting of a red giant (or a Mira-type variable) and 
a hot companion that is usually a white dwarf 
(e.g. Kenyon 1986). Most red
giant components suffer a heavy mass loss in the form of a slow
stellar wind with a typical terminal speed 10 $\sim20 {\rm\ km\ s^{-1}}$
that is comparable to the escape velocity of a giant (e.g. Schmid 1996).
In contrast, many white dwarf systems including planetary nebulae are
known to possess fast outflows with terminal speed 
$\ge 1000\rm\ km\ s^{-1}$. Similar fast winds are also known
in some symbiotic stars including AG~Peg (Vogel \& Nussbaumer 1994).

The orbital elements of symbiotic stars are not well-constrained, but
light curves often show that they possess a long period of several hundred
days (e.g. Iben \& Tutukov 1996). M\"urset \& Schmid (1999) presented the
relation between the spectral types of the cool giants and the orbital
periods to reveal that almost all symbiotic stars are well-detached
binary systems.
Therefore the most important binary activity may be found from the
interaction between the two different kind of winds. These two winds may
collide forming a shocked region, which can be identified with
X-ray observations and is consistent with the fact that
several symbiotic stars are known to be X-ray sources 
(Girard \& Wilson 1987, M\"urset, Wolff \& Jordan 1997,
Ezuka, Ishida \& Makino 1998)

In many symbiotic stars the Raman-scattered 
O~{\tiny{VI}} $\lambda \lambda6827, 7088$ features exhibit 
multiple peak profiles and strong polarization accompanied by 
the polarization flip (Schmid 1989, Schmid \& Schild 1994,
Harries \& Howarth 1996).
Lee \& Park (1999) proposed that these features
can be explained by assuming that there is an accretion disk
around the white dwarf formed through capture of the slow wind from the
giant (e.g. Mastrodemos \& Morris 1998).
According to the theoretical modelling by Paczynski \& Zytkow (1978),
periodic hydrogen shell flashes may occur in a white dwarf with the
accretion rate $10^{-11} - 10^{-7}\rm\ M_{\odot}\ yr^{-1}$, where
each outburst may last for decades. With this eruption the radiative
pressure will drive a stellar wind around a white dwarf, which may
take a bipolar form subject to the circumstellar matter 
distribution \cite{sok2}.
This interpretation is interesting because most symbiotic stars with
known nebular morphologies are bipolar \cite{cor}. 
Currently it is very controversial whether bipolar planetary nebulae
possess central binary systems (e.g. Soker 1998).

It is, therefore, very interesting and important to investigate the
line profiles that indicate the fast outflowing motion, from which
we may find the physical properties associated with the bipolarity
of the wind. Since the resonance doublets arise from
the common
electronic transitions $S_{1/2}-P_{1/2,3/2}$ and have
the separation
ranging from 500${\rm\ km\ s^{-1}}$ (for C~{\tiny{IV}}) to 1,650
${\rm\ km\ s^{-1}}
$ (for O~{\tiny{VI}}), these lines can be an excellent tool to investigate
the outflowing hot wind around the white dwarf component.

When the bulk velocity changes significantly in a 
region that is much smaller than the scale height of the physical
quantities such as density, the radiative transfer
 can be described by the
Sobolev approximation. In this case, the optical depth for a line photon is
inversely proportional to the velocity gradient in the direction of the
photon propagation (e.g. Sobolev 1947; Rybicki \& Hummer 1978).
In the case of resonance doublet
lines, line photons arising from the $S_{1/2}-P_{3/2}$ transition
will be resonantly scattered with $S_{1/2}-P_{1/2}$ transition
by receding ions with the speed of the doublet separation,
if the outflow is an accelerating wind with a speed larger than the  doublet
separation. We denote this type of scattering by `double scattering.'
This double scattering converts line photons associated
with the $S_{1/2}-P_{3/2}$ transition into line photons associated with the
$S_{1/2}-P_{1/2}$ transition. This will change the intrinsic doublet
line ratio, which is 2~:~1 in the optically thin limit and 1~:~1 in the
optically thick limit. Olson (1982) investigated this problem and applied 
to stellar winds around O and B stars. His main concern was limited 
to the investigations of  
various P~Cygni profiles in spherical winds.

\textit{IUE} observations indicate that
many objects including
symbiotic stars and planetary nebulae show various doublet line ratios
between 2~:~1 and 1~:~1 (Feibelman 1983, Schmid et al. 1999).
The microphysical processes that lead to these various line ratios may
be associated with the existence of dust and/or collisional de-excitation
(Ahn \& Lee 2002, in preparation). 

Michalitsianos et al. (1988) showed
that some symbiotic stars including RX Pup and R Aqr show anomalous
line ratios, in which the short wavelength component of the 
C~{\tiny{IV}} doublet
was observed to be weaker than the long wavelength component.
Vogel \& Nussbaumer (1994) showed that the symbiotic nova AG Peg
exhibits broad He~{\tiny{II}} emission lines that are formed in a fast wind.
In this system, the resonance doublets N~{\tiny{V}} and C~{\tiny{IV}} exhibit
inverted line ratios where the short wavelength component is
weaker. 
The authors
proposed that the line formation is strongly affected by the
double scatterings, whereby significant fraction of photons
are converted. Furthermore, the doublets did not display P~Cygni profiles.
These facts imply that the outflowing
motion is confined to specific directions (plausibly in the polar
directions), which exclude the observer's line of sight.

There have been many theoretical investigations on the P~Cygni profiles
shown in the resonance lines of metal elements by adopting the Sobolev
approximation. However, relatively little attention has been paid on 
the profiles observed outside the wind flowing direction. 
In this paper, we perform Sobolev
Monte Carlo computations to obtain the profiles of resonance doublet
lines formed in bipolar winds that may be present in symbiotic stars.

In Section 2, we briefly describe the Sobolev theory and the basic
atomic physics concerning the resonant doublet lines.
We also present the kinematic stellar wind model adopted in this work
and the Monte Carlo procedure.
Our results are presented in Section 3. Finally, we summarise and
discuss our results and observational implications in Section~4.

\section{Wind Model and Calculations}
\subsection{The Sobolev Theory}
In the case of the stellar wind around the white dwarf component in
a symbiotic star, the wind is highly ionised and characterized by
the thermal speed $\sim 10{\rm\ km\ s^{-1}}$ and the
bulk speed of order $10^3{\rm\ km\ s^{-1}}$ (Girard \& Wilson 1987).
We may apply the Sobolev approximation to describe the radiative
transfer,
if the acceleration of the wind material occurs in a small region
where there are no considerable changes in physical quantities. We
will check the validity of the Sobolev approximation later in detail.

When a scattering medium is of moderate column density, wing
scatterings can be safely neglected. In this case, the scattering cross
section associated with a resonance line photon can be
well-approximated by the Dirac $\delta$-function with the strength
${f_{abs}(\pi e^2/m_e c)}$, or
\begin{equation}
\sigma_\nu = f_{abs} \frac{\pi e^2}{m_e c}\delta(\nu-\nu_0),
\end{equation}
where $f_{abs}$ is the absorption oscillator strength associated with
the resonance transition and the frequency $\nu$ is measured in the
rest frame of the scattering atom. Therefore, a scattering occurs
only when the resonance condition is met. Because of the local
thermal motion, the physical width of the scattering region of a
photon is given by the width where the difference of
velocity is of order the thermal velocity $v_{th}$, that is,
$\Delta r \sim v_{th} {(dv/dr)}^{-1}$.

Noting that the local velocity distribution of scattering atoms is
given by the Maxwell-Boltzmann distribution, a straightforward
computation gives the Sobolev optical depth $\tau_{Sob}$ for a given
direction,
\begin{eqnarray}
\lefteqn{\tau_{Sob}=n_i\int \sigma dv
c/[{{\textit{v}}}(d{{\textit{V}}}/d{{\textit{s}}})]}
\nonumber\\
& &=n_if_{abs}\lambda_0\Big(\frac{\pi e^2}{m_e c}\Big)\Big(\frac{dV}
{ds}\Big)^{-1},
\end{eqnarray}
where \textit{s} is the distance along the photon propagation
direction, $\lambda_0$ represents the resonance wavelength, and $n_i$
is the ion number density. A more complete and
detailed description of the Sobolev theory can be found in the
literature (e.g. Sobolev 1947, Rybicki \& Hummer 1978, 
Lee \& Blandford 1997, Ahn, Lee \& Lee 2000).

\begin{figure}
\centerline{ \psfig{figure=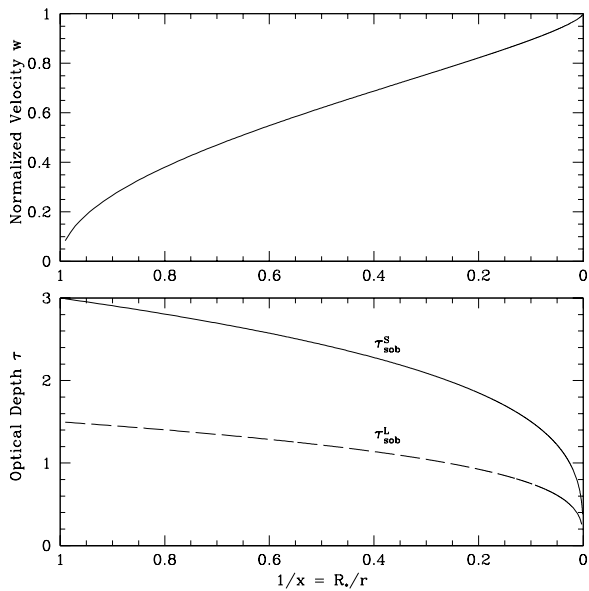,width=8cm}}
\caption[]{The velocity field normalized by the terminal velocity and 
the corresponding Sobolev optical depth $\tau_{Sob}$
according to the distance normalized by $R_*$, the radius of the photosphere
of the hot white dwarf. Solid and dashed lines represent the Sobolev
optical depths of the short and the long wavelength components of the
resonance doublets, respectively}
\label{velocity}
\end{figure}

In this work, we are particularly interested in the resonance
doublets, where the doublet separation is smaller than the wind
velocity. Those doublets include
 C~{\tiny{IV}} $\lambda \lambda1548, 1551$, N~{\tiny{V}} $\lambda \lambda1239,
1243$, and O~{\tiny{VI}} $\lambda \lambda1032, 1038$,
which are known to be prominent in symbiotic stars from
observations made with space instruments such as {\it IUE, the
Hopkins Ultraviolet Telescope (HUT), and the Orbiting and Retrievable 
Far and Extreme Ultraviolet Spectrometer (ORFEUS)}.
The $S_{1/2}-P_{3/2}$ transition corresponding to
the short wavelength component has the twice stronger oscillator
strength than the $S_{1/2}-P_{1/2}$ transition for the long wavelength
component.
Therefore, at the same position, the
Sobolev optical depth $\tau^S_{Sob}$ corresponding to the short
wavelength component is also twice the optical depth
$\tau^L_{Sob}$ for the long wavelength component as are shown in 
Fig.~\ref{velocity}.
From now on, we
refer the Sobolev optical depth $\tau_{Sob}$ to be $\tau^S_{Sob}$
when we do not specify the transition.

\subsection{Kinematics of the Fast Stellar Wind}
The existence of fast stellar winds with speed 
$v\sim 10^3{\rm\ km\ s^{-1}}$ around a white dwarf has been known 
in many white dwarf systems
including the central stars of planetary nebulae
(e.g. Cerruti-Sola \& Petrinotto 1989).
Many P~Cygni profiles formed in fast stellar winds have been successfully
fitted using a velocity law 
$v({\bf{\textit{r}}})=v_\infty (1-r/R_*)^\beta$ with the
Sobolev approximation (Lamers, Cerruti-Sola \& Perinotto 1987), where
the stellar wind is assumed to start at the photospheric radius $R_*$,
$\beta$ is a dimensionless parameter and $v_{\infty}$ is the terminal
velocity.
We scale the physical distance with $R_*$ so that the distance
\textit{r} from the stellar centre is obtained from the
dimensionless parameter $x$ defined by
\begin{equation}
{\bf{\textit{x}}}={\bf{\textit{r}}}/R_*.
\end{equation}
The stellar wind in this work is assumed to be steady and have
geometry in the form of a bipolar cone including the spherically
symmetric case. The velocity {\bf{\textit{v}}}({\bf{\textit{r}}}) of
the wind increases
monotonically outward and asymptotically approaches the terminal
speed \textit{$v_{\infty}$}. Although
the exact model of the velocity law in symbiotic stars is unknown,
Castor, Abbott \& Klein (1975) proposed a very steep velocity
law for a radiation-driven stellar wind. We measure
the wind velocity with the terminal speed, and introduce a dimensionless
velocity $w$ defined by
\begin{equation}
{\bf{\textit{w}}}={\bf{\textit{v}}}/v_\infty.
\end{equation}

Instead of assuming a wind velocity law, we prescribe
the Sobolev optical depth by
\begin{equation}
\tau_{Sob}=\tau_0 \Big(\frac{r}{R_*}\Big)^{-\epsilon}
=\tau_0x^{-\epsilon},
\end{equation}
where $\tau_0$ is the initial Sobolev optical depth at the wind base and 
$\epsilon$ is a dimensionless positive number. We choose $\epsilon$
to be small, so that the Sobolev optical depth does not decrease significantly
in most of the wind region. This will enhance the effect of 
double scattering, where continuum photons blueward of the $S_{1/2}-
P_{3/2}$ transition may get scattered by the transition $S_{1/2}-
P_{1/2}$ further downstream of the wind. We use the initial Sobolev 
optical depth 
$\tau_{Sob,0}^S = 3, \tau_{Sob,0}^L = 1.5$ for the short and the long
wavelength components of the doublet.

The density profile $n({\bf{\textit{r}}})$ is obtained from
the mass flux conservation assuming the constant mass loss rate
\textit{$\dot M$},
\begin{equation}
n({\bf{\textit{r}}})=\frac{\dot M}{4\pi m_p r^2 v({\bf{\textit{r}}})
c_f} \propto \frac{1}{r^2v({\bf{\textit{r}}})},
\end{equation}
where $m_p$ is the proton mass and 
\textit{$c_f$} is the covering factor of the wind 
normalized by the whole sky. From Eqs. (2), (5) and (6),
the velocity law of the wind is given by
\begin{equation}
{\bf{\textit{v}}}({\bf{\textit{r}}})=v_{\infty}\Big[1-\Big(\frac{r}{R_*}
\Big)^{-1+\epsilon}\Big]^{1/2}{\bf{\hat \textit{r}}}.
\end{equation}
In order to check the validity of the Sobolev approximation, we consider
the length scale over which the bulk velocity changes by the amount of
$v_{th}$,
\begin{eqnarray}
L_{Sob} &\equiv& {v_{th}\over dv/dr}  \\
        & =& {2R_*}{v_{th}\over v_\infty}(1-\epsilon)^{-1}
\left(\frac{r}{R_*}\right)^{2-\epsilon}
\left[1-\left(\frac{r}{R_*}\right)^{-1+\epsilon}\right]^{1/2}. \nonumber
\end{eqnarray}
We note that $v_{th}/v_\infty \le 10^{-2}$, and therefore
$L_{Sob}\le 0.02 R_* x^{2-\epsilon}$. This implies that $L_{Sob}$ can
be comparable to $R_*$, when $x\ge 10$ and $\epsilon = 0.3$. When $x$
is small, the density changes steeply near $R_*$, and $L_{Sob}\ll
R_*$ validating the use of the Sobolev approximation. When $x$ gets
larger, the density variation is almost scale-free, which also
makes the Sobolev approximation reliable.

The mass loss rate is related to the Sobolev optical depth
$\tau_{Sob}$ by
\begin{equation}
{\dot M}=5.1\times 10^{-12} \tau_0
(1-\epsilon)c_f\left({v_\infty\over 10^3{\rm\ km\ s^{-1}}}\right)^2
{\rm\ M_\odot}{\rm\ yr^{-1}},
\end{equation}
and the hydrogen number density can be obtained by
\begin{eqnarray}
n({\bf{\textit{r}}})&=&1.6\times10^7\tau_0(1-\epsilon)x^{-2}
w^{-1} \nonumber \\ 
&\times &\left(\frac{R_*}{10^{11}{\rm cm}}\right)^{-2}
\left(\frac{v_\infty}{10^3{\rm\ km\ s^{-1}}}\right){\rm cm}^{-3}.
\end{eqnarray}

According to the study of Cerruti-Sola \& Perinotto (1989),
the terminal velocities of the fast winds in many planetary nebulae
range $1000-2500{\rm\ km\ s^{-1}}$. Vogel \& Nussbaumer (1994) 
also showed that
the fast wind in the symbiotic star AG~Peg is about 1000${\rm\ km\
s^{-1}}$. In this study, we choose the doublet separation to be
$960{\rm\ km\ s^{-1}}$ corresponding to the N~{\tiny{V}} doublet
and set the terminal velocity
$v_\infty= 3000{\rm\ km\ s^{-1}}$ in order to investigate the 
effect of double scattering. Therefore, we focus on the profiles
formed by various scatterings, not on the full dynamical range of the wind.

With the choices of
the parameters $\epsilon=0.3$, $\tau_0=3$ and the wind half opening
angle $30\degr$, the mass loss rate 
${\dot M}= 1.3\times 10^{-11} {\rm\ M_\odot}{\rm\ yr^{-1}}$. 
It is notable that the mass loss rate is compatible to those of 
Michalitsianos et al. (1988) which are
$4.5\times10^{-12} \le  \dot M \le  1.0\times10^{-11} {\rm\ M_{\odot}}
{\rm\ yr^{-1}}$ for R~Aqr and 
$1.5\times10^{-11} \le  \dot M \le  3.1\times10^{-11} {\rm\ M_{\odot}}
{\rm\ yr^{-1}}$
for RX~Pup. According to them, it is estimated that the momentum flux from 
the red giant
is about 200 times greater than that from the hot companion. 
The corresponding mass loss
rate from the red giant is $1.3\times10^{-7} {\rm\ M_{\odot}}{\rm\ yr^{-1}}$
which is consistent with the calculation by Vogel \& Nussbaumer (1994).

The profiles of the wind velocity and the Sobolev optical depth
are shown in Fig.~\ref{velocity} as a function of the normalized
distance $x$.

\begin{figure}
\centerline{ \psfig{figure=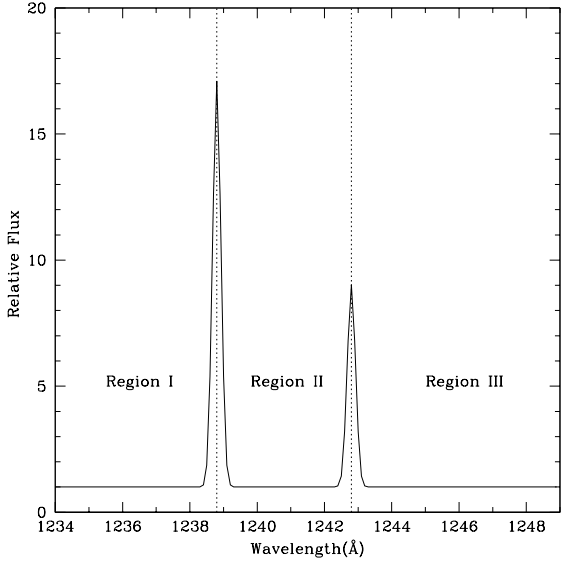,width=8cm}}
\caption[]{The spectrum of the source that consists of a flat continuum
and resonance doublet lines, which we assume is N~{\tiny{V}} here.
We used Gaussian profiles for the line doublet whose equivalent widths
are 5$\rm\ \AA$ and $2.5\rm\ \AA$ for the short and the long wavelength 
components, respectively. The velocity dispersion is chosen to be
$\sigma=30{\rm\ km\ s^{-1}}$ for both components. The
wavelength regions are divided according to the wavelength relative to
the resonance line centres of the doublet.}
\label{region}
\end{figure}

\begin{figure}
\centerline{ \psfig{figure=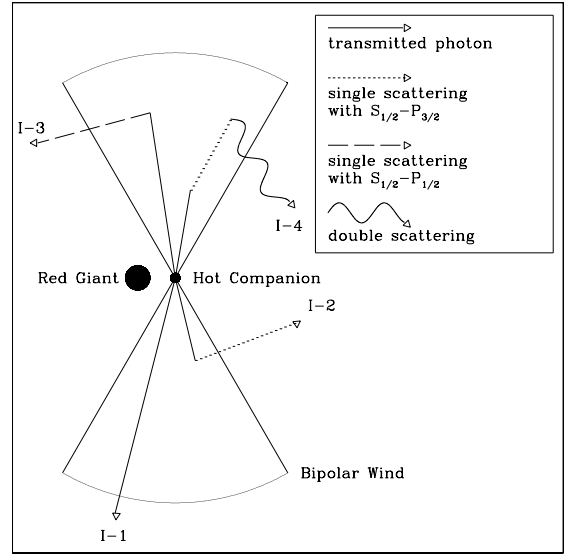,width=8cm}}
\caption[]{Conceptual diagram for single resonance scatterings
and double resonance scatterings.
The solid, dotted, dashed and wavy lines represent no scattering,
single scattering resonant with $S_{1/2}-P_{3/2}$, 
single scattering resonant with $S_{1/2}-P_{1/2}$
and double scattering, respectively. The indices represent
possible scatterings that a given photon in Region~I can suffer.}
\label{photon}
\end{figure}


\subsection{The Sobolev Monte Carlo Code}
Firstly, we consider the volume emission measure in the stellar wind 
region. The 
relevant quantity to be considered is 
\begin{eqnarray}
\lefteqn{\int n({\bf{\textit{r}}})^2 dV
=n_0^2R_*^3\tau_0^2(1-\epsilon)^24\pi \int_1^\infty
\frac{dx}{x^2-x^{1+\epsilon}}} \nonumber \\
&\verb+      + &=n_0^2R_*^3\tau_0^2(1-\epsilon)^24\pi f(\epsilon),
\end{eqnarray}
where $n_0=1.6\times10^7 \rm\ cm^{-3}$ and
$f(\epsilon) \equiv \int_1^\infty \frac{dx}{x^2-x^{1+\epsilon}}$
which is a steep function of $\epsilon$ near zero.
For a small value $\epsilon =0.3$, we also have a moderate value of
$f(\epsilon)$ and the quantity $n^2V$ is basically determined by
$n_0^2R_*^3$ $\simeq 10^{47}\rm\ cm^{-3}$ much smaller than typically
observed in symbiotic stars (e.g. Proga, Kenyon \& Raymond 1998).
Therefore, if we consider the wind as a photon 
scattering region, almost all the photons are generated 
near the hot star component.
This means that the photon source is separated from the wind.

From this consideration, we prepare a photon source near the white
dwarf component, from which both continuum photons and line photons
are injected into the wind region. We choose the equivalent width of 
the short wavelength component to be $5\rm\ \AA$ and $2.5\rm\ \AA$ for
the long wavelength component. This ratio corresponds to the optically 
thin limit of the emission plasma. Without specifying the details
of the kinematics of the emission region, we just assume that
the line profile is given by 
a Gaussian with the velocity dispersion $\sigma =30 \rm\ km\ s^{-1}$
for both components. Fig.~\ref{region} shows the profile of the 
incident photon flux that is injected into the wind region.

According to the wavelength relative to the two resonance line centres,
we conceptually divide the injected photons into three categories, which
we call Region I, II and III shown in Fig.~\ref{region}.

For photons in Region~III, there is no point in the wind where the
resonance condition can be satisfied for either transition.
Therefore, no photons in Region~III will be scattered in the wind in our
model. Photons from Region~II can only be resonantly scattered in the wind
with the $S_{1/2}-P_{1/2}$ transition, for which the relevant Sobolev
optical depth is $0.5\tau_{Sob}$. 

Most complicated interactions can be seen
for photons in Region~I, because they can be first resonantly scattered
with the $S_{1/2}-P_{3/2}$ transition and can be additionally scattered
with the $S_{1/2}-P_{1/2}$ transition. 
Hence there are four cases that can occur for photons in Region~I. The first
case denoted by I-1 in Fig.~\ref{photon} corresponds to the escape 
without any scattering. 
The second case (I-2) represents the escape after a resonance scattering
associated with $S_{1/2}-P_{3/2}$ transition. The third case (I-3)
corresponds to the escape after a resonance scattering associated with
$S_{1/2}-P_{1/2}$ transition. The final case (I-4) is the double scattering,
where a given photon escapes from the bipolar wind after resonance 
scatterings associated with both transitions. Fig.~\ref{photon} illustrates
these four possible cases.

\begin{figure}
\centerline{ \psfig{figure=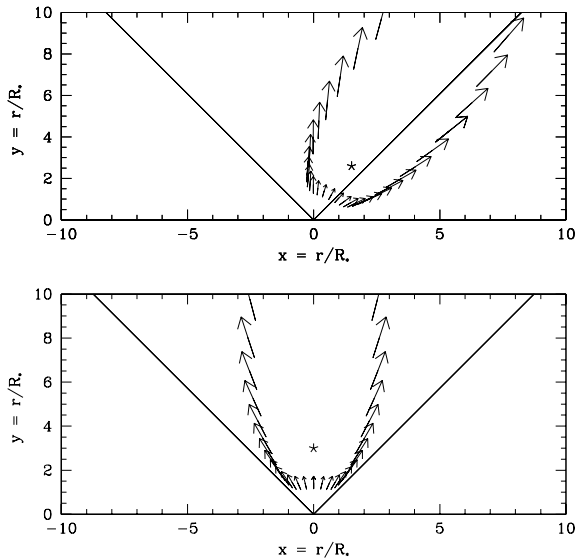,width=8cm}}
\caption[]{Examples of the Sobolev surfaces for double scattering 
in the $x-y$ plane.
A given photon is resonantly scattered at the position marked by an asterisk,
and the surfaces with the velocity component equal to the doublet separation.
The arrows and the solid lines represent the velocity vectors at the
positions and the wind region, respectively.}
\label{double}
\end{figure}
\begin{figure}
\centerline{ \psfig{figure=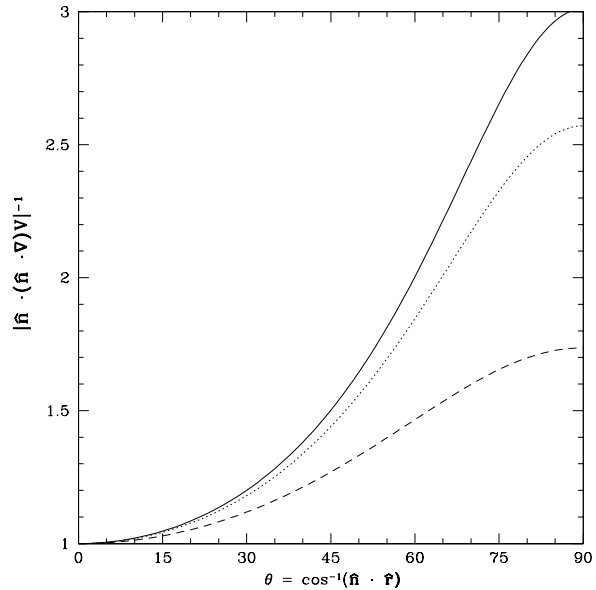,width=8cm}}
\caption[]{The dependence of the Sobolev optical depth $\tau_{Sob}$ on the
photon propagation direction at a given position. The horizontal axis 
represents the angle between the incident photon wavevector and the radial
or the wind direction. The solid, dotted and dashed
lines represent the inverse of the velocity gradient at
x=1.16, 1.2 and 1.3 which are normalized by the inverse value of the
velocity gradient for the radial direction.}
\label{sobolev}
\end{figure}

We consider the Sobolev surface, which is defined as the collection of
those points moving away from a given point with the same velocity component
along the line of the photon propagation. In order to describe doubly scattered
photons, we have to consider the Sobolev surface moving away with the velocity
of the doublet separation from a given point where the first resonance
scattering occurs. For illustration, in Fig.~\ref{double} we show a couple
of Sobolev surfaces for double scattering in $x-y$ plane, where the given
photon is firstly resonantly scattered at the positions marked by an 
asterisk. The Sobolev optical depth $\tau_{Sob}^D$ associated with this 
double scattering is given by

\begin{eqnarray}
\lefteqn{\tau_{Sob}^D=n_if_{abs}^L\lambda_0\Big(\frac{\pi e^2}{m_e c}\Big)
|\mbox{\boldmath $\hat n \cdot(\hat n \cdot \nabla)V$}|^{-1}}
\nonumber\\
\lefteqn{=n_if_{abs}^L\lambda_0\Big(\frac{\pi e^2}{m_e c}\Big)
\Big|\mbox{(\boldmath $\hat n \cdot \hat r$})^2
\frac{dV}{dr}+\frac{V}{r}[1-\mbox{\boldmath $(\hat n \cdot \hat r$})^2
]\Big|^{-1},}
\end{eqnarray}
where {\bf $\hat \textit{n}$} is the photon wave vector before double 
scattering and $f_{abs}^L$ is the oscillator strength corresponding to
the $S_{1/2}-P_{1/2}$ transition.

In Fig.~\ref{sobolev}, we show the inverse of the velocity gradient for
various directions at three positions $x=1.16, 1.2, 1.3$ in order
to find the dependence of $\tau_{Sob}^D$ on the photon propagation
direction. We can see that $\tau_{Sob}^D$ may vary by a factor of three
dependent on the direction.

Since we are particularly interested in bipolar winds, which may be related
with the bipolar nebular morphologies known in a number of symbiotic stars
(e.g. Corradi 1995, Corradi \& Schwarz 1993),
it is natural to expect that there exists
a very thick circumstellar component in the equatorial plane. Therefore,
in our model, we introduce an opaque medium outside the photon source and
the wind base. For simplicity, we set the physical size of the opaque medium
to be $x=1.01$, so that observer in the polar direction can see the wind
base, which is hidden from the equatorial direction.

We typically inject $10^5$ continuum photons in a bin with 
$\Delta \lambda = 0.1\rm\ \AA$. The injection number is adjusted near
the doublet line centres so that the line features possess the
equivalent widths of 5 \AA\ and 2.5 \AA\ and the Gaussian profiles 
with velocity dispersion $\sigma= 30{\rm\ km\ s^{-1}}$. 
We collect emergent photons according to
the direction cosine of the wave vector with a bin size of 
$\Delta \mu = 0.1$, and the wavelength bin $\Delta \lambda = 0.023\rm\ \AA$
corresponding to the thermal width of $ T=10^4\rm\ K$. In our simulations, we
normalize the relevant quantities to those associated with N~{\tiny{V}}
$\lambda \lambda 1238.8, 1242.8$.

\section{Results}
We investigate the dependence of the profiles on the observer's
line of sight, the mass loss rate,
the wind half opening angle and the existence of the opaque
circumstellar region.

\begin{figure*}
\centerline{ \psfig{figure=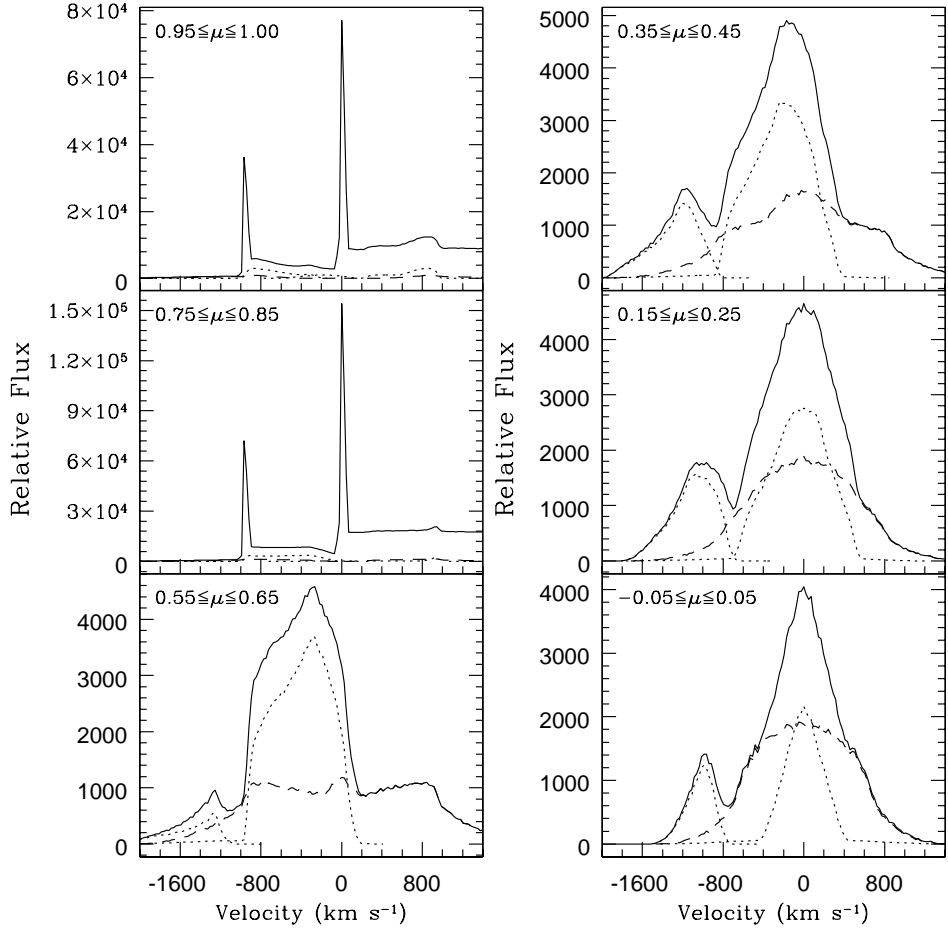,width=13cm}}
\caption[]{Profiles according to the various observer's lines of
sight. 
The number of photons counted in each observer's line of
sight is referred to the relative flux and is shown as a solid
line. The dotted and dashed lines mean the number of photons 
which suffered single scattering and the double scattering,
respectively. The wind
half opening angle is set to be $45\degr$.}
\label{result1}
\end{figure*}

\begin{figure*}
\centerline{ \psfig{figure=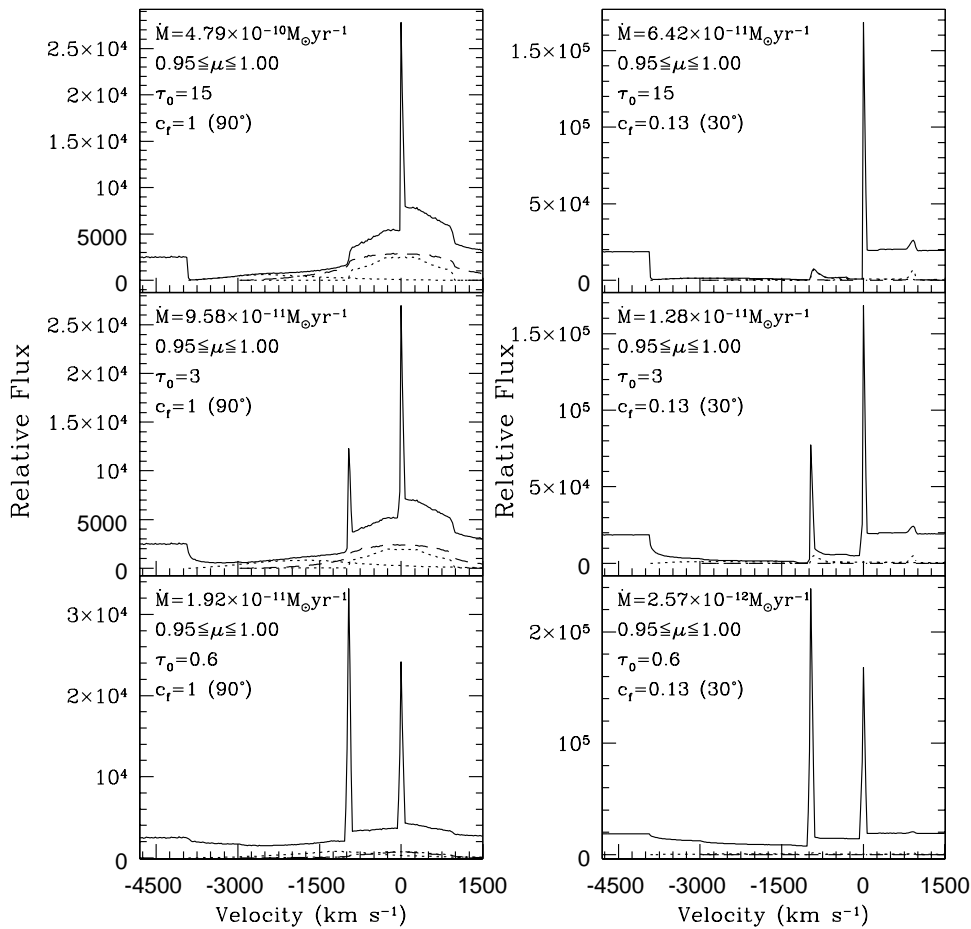,width=13cm}}
\caption[]{Profiles for various mass loss rates ${\dot M}$.
We fix the observer's line of sight to be in the polar direction.
The profiles shown in the left correspond to spherical winds.
In the right part are shown
the profiles for bipolar winds with the half opening angle 30\degr.
The mass loss rates are chosen so that the initial Sobolev optical
depth $\tau_0=0.6, 3$ and 15 and $c_f$ represents the covering
factor of the wind with respect to the photon source.
The various lines represent the same quantities illustrated in 
Fig.~\ref{result1}}
\label{result1_1}
\end{figure*}

\subsection{Profiles for Various Lines of Sight}
We first fix the wind half opening angle to be 45$\degr$ and
consider the variation of the profiles dependent on the
observer's line of sight. Fig.~\ref{result1} shows
our result, in which we find significant difference in the profiles
viewed from the polar direction and from the equatorial direction.

When $\tau_{Sob}$ is much larger than 1 and the observer is in the polar  
direction, we normally obtain P~Cygni type profiles. However,  when the 
Sobolev optical depth  $\tau_{Sob}$ is moderate  ($\tau_{Sob}\simeq 1$), 
the observer sees profiles characterized by two narrow peaks. The blue 
part of the short wavelength peak is obtained from suppression of 
the incident flux  by a  factor $e^{-\tau_{Sob}(r_1)-0.5\tau_{Sob}(r_2)}$ 
with 
$r_1$ and $r_2$ being the locations of the resonance transitions with 
$S_{1/2}-P_{3/2}$ and $S_{1/2}-P_{1/2}$, respectively. In contrast, 
the red 
part of the short wavelength peak and the blue part of the long wavelength 
peak are suppressed by a factor $e^{-0.5\tau_{Sob}(r_3)}$, where  $r_3$ is 
the location of the resonance scattering with $S_{1/2}-P_{1/2}$.
Therefore the double scattering process affects the doublet line ratio
significantly.
In our model, only the blue part of the long wavelength component can
satisfy the resonance condition and be affected by the wind.
Hence we observe the fairly
stronger long wavelength component than the short wavelength.
This is interesting considering the \textit{IUE} spectra of RX~Pup and R~Aqr
which were pointed out by Michalitsianos et al. (1988).
According to them, in RX~PUP and R~Aqr the C~{\tiny{IV}} doublet
showed anomalous line ratio in the blue part only, where the short
wavelength component is weaker than the long wavelength component.
However, in the red part, the doublet flux ratio almost restores the 
optically thick limit of 1:1. Therefore, it is very probable that the
doublet line ratio in the blue part may be affected by extrinsic 
components including the fast outflowing wind.

As is apparent from Fig.~\ref{result1}, the continuum blueward
of the short wavelength component (Region I in Fig.~\ref{region})
is weaker than that for Region II, which is again weaker
than that for Region III. A similar trend
is observed for $\mu \ge 0.85$ (in the direction of the wind region).
However, qualitatively different
profiles are obtained for $\mu \le 0.65$, for which the observer is
outside the wind flowing direction.

If we see symbiotic stars outside the wind flowing direction 
($\mu \le 0.65$ in Fig.~\ref{result1}), the emergent profiles 
are composed of only
scattered photons and
no transmitted continuum can be seen. One of the most interesting features
of this bipolar wind system is shown for $\mu \simeq 0$, where we see the
clear effect of the photon redistribution by the double scattering process.
The net effect of double scattering is to take out the flux of
the short wavelength component and redistribute it around the long wavelength
component. Therefore, the flux around the long wavelength component
increases. 

More quantitatively, the ratio of the fluxes contributed by
single scatterings at the short and the long wavelength components and
double scattering to the total flux
is given by 15\%, 39\% and 46\% in the particular case $0.15 \le \mu \le 0.25$
in Fig.~\ref{result1}.
This means that more than half of 1239 photons are doubly
scattered and converted into 1243 photons. 
It is notable that the profile around
the long wavelength component gets greater dispersion
than the dispersion that would be obtained in the case 
when only  single scattering at the long 
wavelength component is operational. This is 
because doubly scattered photons acquire the Doppler shift 
corresponding to the doublet separation
in addition to the Doppler shift for the velocity at the scattering point. 
Therefore, the doublets in a bipolar wind are characterized by
the inverted flux ratio and much
broadened profiles of the long wavelength component.

\begin{figure*}
\centerline{ \psfig{figure=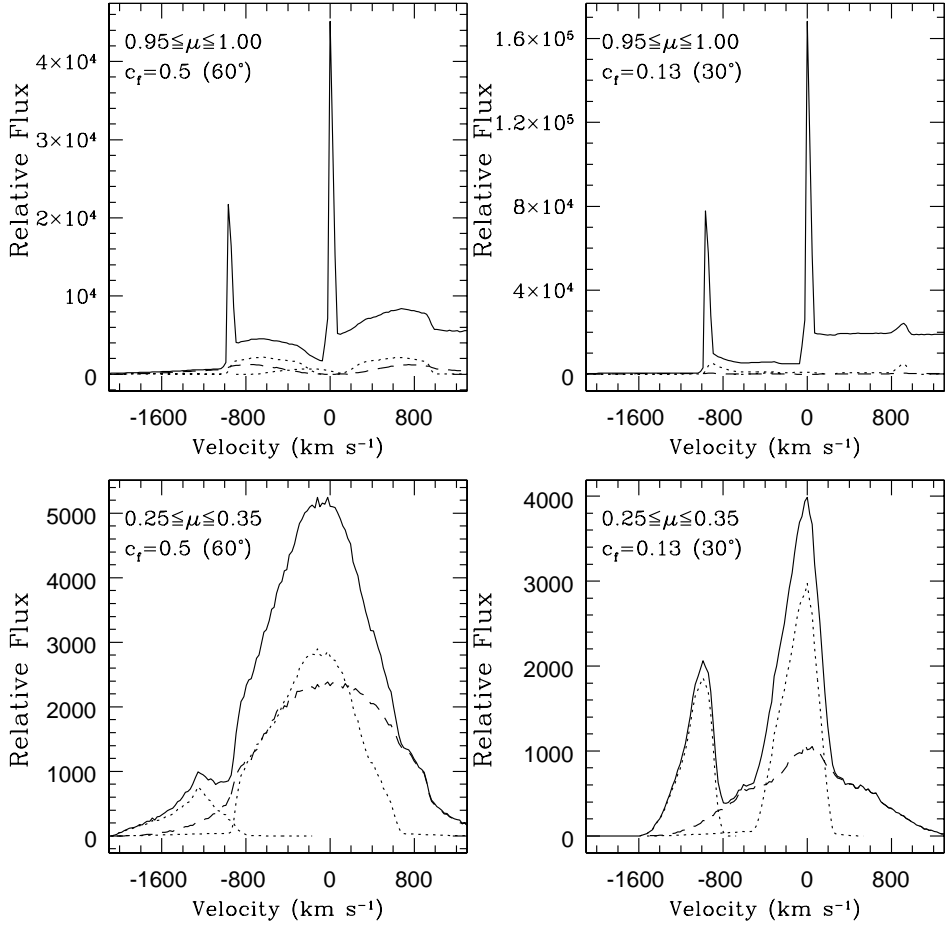,width=13cm}}
\caption[]{The profiles for the wind with half opening angle $60\degr$
and $30\degr$.
The various lines represent the same quantities illustrated in 
Fig.~\ref{result1}}
\label{2-1}
\end{figure*}

\subsection{Dependence on the Mass Loss Rate}
In order to investigate the dependence of the profiles on the mass loss
rate $\dot M$, we set the covering factor of the wind to be 1 and 0.13 
corresponding to the spherical and bipolar winds with
half opening angles 90\degr and 30\degr, respectively.
In this subsection, we fix the observer's line of sight to be 
the polar direction. 

According to the various mass loss rates, considerably different 
profiles are obtained as shown in Fig.~\ref{result1_1}. The mass loss rate
is proportional to the initial optical depth $\tau_0$ as in Eq.~(9).
Various doublet line ratios are obtained because
the peaks of the resonance doublet are sensitive to $\tau_0$.
For the spherical wind, the Sobolev surface for double scattering at
a given position subtends a fairly large solid angle, and therefore
the emergent flux profile will be significantly affected by photons
redistributed by double scattering. In particular, when the mass
loss rate is very high (the top-left panel of Fig.~\ref{result1_1}),
the suppression of the short wavelength component is so large that
no apparent line feature is found around at $1238.8\rm\ \AA$. 
Furthermore, all the photons blueward of the $S_{1/2}-P_{3/2}$ resonance
are redistributed in frequency space up to the wind terminal speed and
form the strong and broad component around the $S_{1/2}-P_{1/2}$ 
resonance wavelength.

However, for the narrow bipolar winds,
the Sobolev surface for double scattering has a small solid angle, 
where we expect a negligible contribution from doubly scattered photons.
Since the scattering process in the bipolar wind
can be effectively regarded as a photon annihilation process
from the observer's line of sight, we see a quick disappearance of the 
broad components around the doublet formed by the photons redistributed
by single scattering.

\subsection{Dependence on the Wind Opening Angle}

We briefly investigate the profile dependence on the wind opening
angle in order to study the profile formation in a bipolar wind.
As a reference, the profile from a spherical wind can be found 
in Fig.\ref{result1_1}.
The inverted intensity ratio will be sensitively dependent on the mass
loss rate. Around the line centres, there
appear broad bases which are formed by those photons redistributed
from single resonant scatterings associated with
$S_{1/2}-P_{1/2,3/2}$ transitions, and also contributed from
double resonant scatterings of photons initially generated in 
Region~I.

In Figs.~\ref{2-1}, we generated profiles for
bipolar winds with half opening angles $60\degr$ and $30\degr$,
respectively. We chose two lines of sight for illustration.
The overall trend is explained in the previous subsection. As the
half opening angle decreases, the redistributed photons contribute less
to the profile formation. This effect is clearly seen in the case
$0.25 \le \mu \le 0.35$, for example. In a bipolar wind with $60\degr$
half opening angle,
we can barely recognise the short wavelength component, whereas the
long counterpart is enhanced significantly. In this case, there is
effectively one broad line component that is centred nearly at
$1242.8\rm\ \AA$, slightly blueward of the long wavelength
component. However,
in the case of the $30\degr$ bipolar wind, there appear two
clear line components of which the longer one is the stronger.
The contribution of redistributed photons is significantly
suppressed. The long wavelength component is, however, broader
than the short wavelength component and there is also a broad base
redward of the long wavelength component.

The fluxes observed from the equatorial direction display various
profiles that are not of a P~Cygni type.
These various profiles are typified by inverted line ratios,
and are formed by photons redistributed by single
and/or double scatterings, which play an essential role in profile
formation. 

\subsection{Miscellaneous Cases}

\begin{figure*}
\centerline{ \psfig{figure=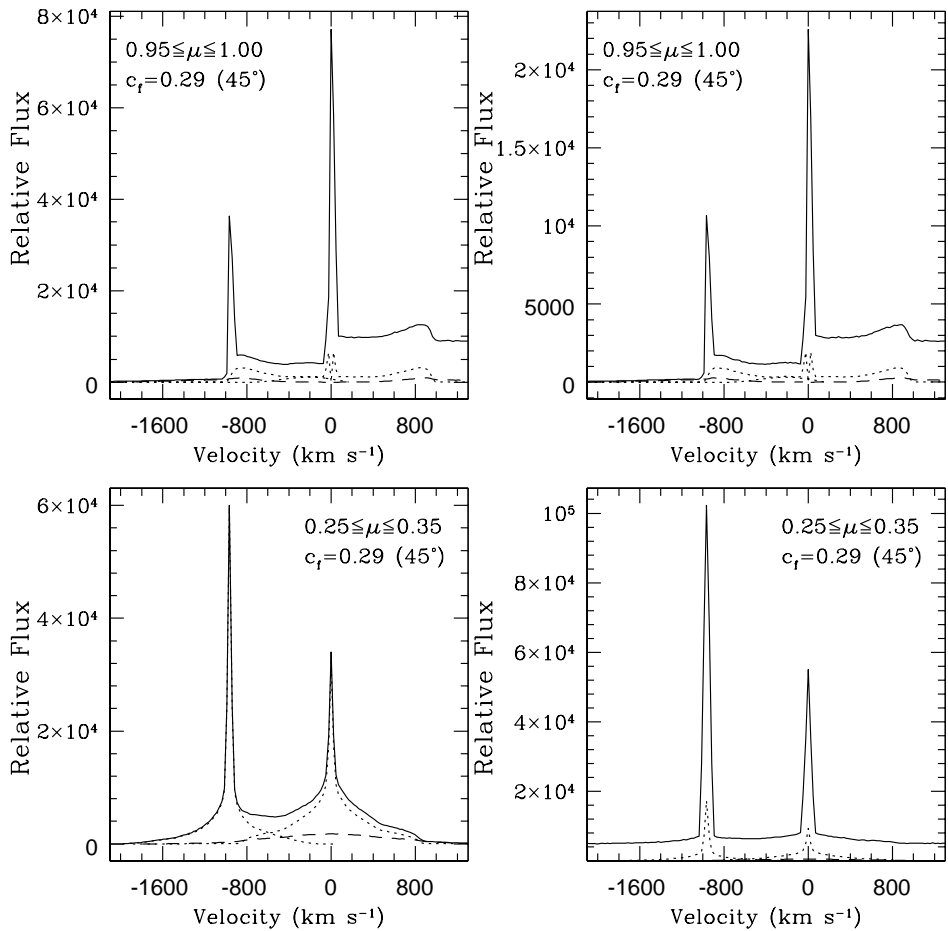,width=13cm}}
\caption[]{Profiles for the cases where the opaque media are changed.
In the left, the opaque media are shrunken to hide only the photon 
injecting source and the wind base is fully seen from all directions.
The opaque media are completely removed in the right, so that
the photon injecting source and the wind base are fully seen.
The wind half opening angle is set to be $45\degr$.
The various lines represent the same quantities illustrated in 
Fig.~\ref{result1}}
\label{no}
\end{figure*}

In Fig.~\ref{no}, we shrink the opaque component, so that
it only hides the photon source. By doing this, effectively we
consider the case where the wind base part is fully seen from any
directions.

When an observer is in the polar direction, we get the same result 
as before. However,
from the equatorial direction we obtain different profiles
to which the wind base part contributes additionally with very small
Doppler shifts. For the equatorial viewer, the wind base part
actually acts as a narrow line source which is almost at
rest. Therefore, the qualitative difference is due to the addition of the
narrow line component to the broad scattered profiles.

Finally, when we remove the opaque component (Fig.~\ref{no}), then
we see fully the source and the scattering wind. No discernible
difference is found for profiles observed from the polar direction.
For an observer in the equatorial direction, the profile is dominantly
that of the source and modified in a very small amount by scattered
components.

\section{Discussion and Summary}
In this paper we investigate the doublet resonance line formation 
in bipolar conic flows in symbiotic stars by adopting the Sobolev Monte
Carlo method. We also pay special attention to the doublet line
ratios.

When an observer is in the polar direction, there appear two peaks at
the line centres coinciding with those of the resonance doublets.
This is because we chose the mass loss rate $\dot M$ so that the Sobolev
optical depth $\tau_{Sob} \simeq 1$ for most of the velocity space,
and therefore the line flux of the short component is suppressed by
a factor $e^{-\tau_{Sob}} \simeq 0.3$. 
The red part
of the long wavelength peak does not suffer any suppression. This fact
implies that 
the detailed peak profiles differ in a systematic way. The suppression pattern 
also repeats for continuum parts, where the weakest occurs blueward of the 
short wavelength peak.

Michalitsianos et al. (1988) investigated the profiles of C~{\tiny{IV}} 
$\lambda \lambda 1548, 1551$ in 
the symbiotic stars RX~Pup and R~Aqr, from which they found the weaker short 
wavelength component than the long wavelength component. They particularly 
noted that the profiles are characterized by 
significant suppression 
in the blue part of the short wavelength peak. 
The peak flux ratio is  most sensitively dependent on  the Sobolev optical 
depth, particularly when $\tau_{Sob}\simeq 1$ and is insensitive on the 
covering factor of the wind.

In RX~Pup and R~Aqr, there may exist a scattering component outflowing 
with speed less than the doublet separation. In this case,
only single scattering is operational, which can cause
the deviations of the doublet line ratio from the optically thin limit. 
However, when double 
scattering is possible, much more various doublet line ratios can be produced. 
It is uncertain that there exists an outflow moving faster than the 
doublet separation of 500${\rm\ km\ s^{-1}}$ in RX~Pup and R~Aqr.
In order to find the concrete 
evidence of double scattering, we need to obtain
spectra with sufficient quality to discern  the continuum level which 
are contributed by the redistributed photons.


When we observe in the equatorial direction, we find that the profiles 
are characterized by two broad components, where the long wavelength 
component is the broader and stronger of the two. This qualitative feature
does not change even if we vary either the mass loss rate or 
the Sobolev optical depth $\tau_{Sob}$ as long as $\tau_{Sob}$ exceeds unity. 
However, when the observer is in the polar direction, various P~Cygni
type profiles are obtained dependent on the mass loss rate.
We believe that the 
bipolarity of the stellar wind is mainly responsible for these spectroscopic
characteristics. 

It is quite interesting that the symbiotic nova AG Peg
shows the profiles of N~{\tiny{V}} and C~{\tiny{IV}} in the hot
wind phases similar to those 
obtained in this work. We strongly believe that the hot wind in AG Peg
takes the bipolar form with the speed exceeding $10^3{\rm\ km\ s^{-1}}$.
Kenny, Taylor \& Seaquist (1991) used the Very Large Array (VLA) in order
to investigate the morphological structure of AG Peg. They found a bipolar
outer nebula extending $\sim 40{\rm\ ''}$. They also found a bipolar
enhancement in the inner nebula having a subarcsecond extent, which is
consistent with the existence of a bipolar wind. However, it should also
be noted that Nussbaumer,
Schmutz \& Vogel (1995) reported a P~Cygni type absorption in
N~{\tiny{V}} in AG~Peg
using the {\it Hubble Space Telescope}.  

Most symbiotic stars are not  optically
resolved and therefore their morphology is very uncertain. 
In planetary nebulae, it is still highly controversial 
whether the binarity of the central star system can be linked to
the bipolarity of the nebula (e.g. Morris 1987, Livio \& Soker 1988,
Soker 1998).
A significant input should be provided from the morphological study of 
symbiotic stars. In this respect, it is proposed that 
the resonance doublets can be 
a useful diagnostic of the bipolar winds in symbiotic stars.

\section*{Acknowledgments}
J.J.Y. is grateful to Chan-Gyung Park for useful discussions and indebted
to Myungshin Im for encouragement for this research work.
This work was supported by Korea Research Foundation Grant 
(KRF-2001-003-D00105).
\bibliographystyle{mnras}
\bibliography{reference}

\begin{thebibliography}{}
\bibitem[\protect\citefmt{Lee et al}{2002}]{lea}
Ahn S.-H., Lee H.-W., 2002, in preparation.

\bibitem[\protect\citefmt{Lee et al}{2000}]{all}
Ahn S.-H., Lee H.-W., Lee H. M., 2000, Journal of the Korean 
Astronomical Society, 33, 29

\bibitem[\protect\citefmt{Castor, Abbott, Klein}{1975}]{cak}
Castor J. I., Abbott D. C., Klein R. I., 1975, ApJ, 195, 157

\bibitem[\protect\citefmt{Corradi}{1995}]{cor}
Corradi R., 1995, MNRAS, 276, 521

\bibitem[\protect\citefmt{Corradi, schwarz}{1995}]{cns}
Corradi R., Schwarz H., 1993, A\&A, 267, 714

\bibitem[\protect\citefmt{Cerruti-Sola}{1989}]{cer}
Cerruti-Sola M., Perinotto M., 1989, ApJ, 345, 339

\bibitem[\protect\citefmt{Ezuka}{1998}]{eim}
Ezuka H., Ishida M., Makino F., 1998, ApJ, 499, 388

\bibitem[\protect\citefmt{Feibelman}{1983}]{fei}
Feibelman W. A., 1983, A\&A, 122, 335

\bibitem[\protect\citefmt{Girard et al}{1987}]{gir}
Girard T., Wilson L.A., 1987, A\&A, 183, 247

\bibitem[\protect\citefmt{Harries and Howarth}{1996}]{hah}
Harries T. J., Howarth I. D., 1996, A\&AS, 119, 61

\bibitem[\protect\citefmt{Iben}{1996}]{ibe}
Iben I. J., Tutukov A. V., 1996, ApJS, 105, 145

\bibitem[\protect\citefmt{Kenny}{1991}]{kenn}
Kenny H. T., Taylor A. R., Seaquist E. R., 1991, ApJ, 366, 549

\bibitem[\protect\citefmt{Kenyon}{1986}]{ken}
Kenyon S. J., 1986, The Symbiotic Stars., Cambridge Univ. Press, Cambridge

\bibitem[\protect\citefmt{lamers}{1987}]{lam}
Lamers H. J. G. L. M., Cerruti-Sola M., Perinotto M., 1987, ApJ, 
314, 726

\bibitem[\protect\citefmt{Lee and Park}{1999}]{lnp}
Lee H.-W., Park M.-G., 1999, ApJ, 515, L89

\bibitem[\protect\citefmt{Lee}{1997}]{lnb}
Lee H.-W., Blandford R. D., 1997, MNRAS, 288, 19

\bibitem[\protect\citefmt{Livio and Soker}{1988}]{las}
Livio M., Soker N., 1988, ApJ, 329, 764

\bibitem[\protect\citefmt{Mastrodemos and Morris}{1998}]{mam}
Mastrodemos N., Morris M., 1998, ApJ, 497, 303


\bibitem[\protect\citefmt{Michalitsianos et al}{1988}]{mic}
Michalitsianos A. G., Kafatos M., Fahey R. P., Viotti R., Cassatella A.,
Altamor A., 1988, ApJ, 331, 477

\bibitem[\protect\citefmt{Morris}{1987}]{mor}
Morris M., 1987, PASP, 99, 1115

\bibitem[\protect\citefmt{Muerset}{1997}]{mue}
M\"urset U., Wolff B., Jordan S., 1997, A\&A, 319, 210

\bibitem[\protect\citefmt{Muerset}{1999}]{mue2}
M\"urset U., Schmid H. M., 1999, A\&AS, 137, 473

\bibitem[\protect\citefmt{Nussbaumer}{1995}]{nus}
Nussbaumer H., Schmutz W., Vogel H., 1995, A\&A, 293, L13

\bibitem[\protect\citefmt{Olson}{1982}]{ols}
Olson G. L., 1982, ApJ, 255, 267

\bibitem[\protect\citefmt{Paczynski \& Zytkow}{1978}]{pac}
Paczynski B., Zytkow A., 1978, ApJ, 222, 604

\bibitem[\protect\citefmt{Proga}{1998}]{pro}
Proga D., Kenyon S., Raymond J., 1998, ApJ, 501, 339

\bibitem[\protect\citefmt{Rybicki}{1978}]{ryb} 
Rybicki G. B., Hummer D. G., 1978, ApJ, 219, 654

\bibitem[\protect\citefmt{Schmid}{1989}]{sc} 
Schmid H. M., 1989, A\&A, 211, 31

\bibitem[\protect\citefmt{Schmid}{1996}]{sc2} 
Schmid H. M., 1996, MNRAS, 282, 511

\bibitem[\protect\citefmt{Schmid}{1999}]{sch} 
Schmid H. M. et al., 1999, A\&A, 348, 950

\bibitem[\protect\citefmt{Schmid}{1994}]{schm} 
Schmid H. M., \& Schild H., 1994, A\&A, 281, 145

\bibitem[\protect\citefmt{Sobolev}{1947}]{sob} Sobolev V. V., 1947,
Moving Envelopes of Stars. Leningrad State Univ., Leningrad (English 
translation: 1960, Gaposchkin S., Havard Univ. Press, Cambridge, MA)

\bibitem[\protect\citefmt{Soker}{1998}]{sok}
Soker N., 1998, ApJ, 496, 833

\bibitem[\protect\citefmt{Soker \& Rapparport}{2000}]{sok2}
Soker N., Rappaport S., 2000, ApJ, 538, 241

\bibitem[\protect\citefmt{Vogel and Nussbaumer}{1994}]{vnn}
Vogel M., Nussbaumer H., 1994, A\&A, 284, 145


\end{thebibliography}

\end{document}